\documentclass[useAMS,usenatbib]{mn2e}
\usepackage{graphicx}
\usepackage{natbib}
\usepackage{amsmath,amssymb}
\usepackage{bbm}
\usepackage{hyperref}
\hypersetup{
   colorlinks=true,
   urlcolor=blue,
   citecolor=blue,
   pdfborder= 0 0 0
}

\title[Satellite alignment in cosmic web]{The alignment of satellite galaxies and cosmic filaments: observations and simulations}
\author[E.~Tempel, Q.~Guo, R.~Kipper, and N.~I.~Libeskind]
{E.~Tempel$^{1}$\thanks{E-mail: elmo.tempel@to.ee},
Q.~Guo$^{2}$, R.~Kipper$^{1,3}$, and N.~I.~Libeskind$^{2}$\\
$^{1}$Tartu Observatory, Observatooriumi~1, 61602 T\~oravere, Estonia\\
$^{2}$Leibniz-Institut f\"ur Astrophysik Potsdam, An der Sternwarte 16, D-14482 Potsdam, Germany\\ 
$^{3}$Institute of Physics, University of Tartu, Ravila 14c, 51010 Tartu, Estonia}

\begin{document}

\date{Accepted 2015 April 22.  Received 2015 April 22; in original form 2015 February 9}

\pagerange{\pageref{firstpage}--\pageref{lastpage}} \pubyear{2015}

\maketitle

\label{firstpage}

\begin{abstract}

{The accretion of satellites onto central galaxies along vast cosmic filaments is an apparent outcome of the anisotropic collapse of structure in our Universe. Numerical work (based on gravitational dynamics of $N$-body simulations) indicates that satellites are beamed towards hosts along preferred directions imprinted by the velocity shear field. Here we use the Sloan Digital Sky Survey to observationally test this claim. We construct 3D filaments and sheets and examine the relative position of satellite galaxies. A statistically significant alignment between satellite galaxy position and filament axis in observations is confirmed. We find a qualitatively compatible alignments by examining satellites and filaments similarly identified in the Millennium simulation, semi-analytical galaxy catalogue. We also examine the dependence of the alignment strength on galaxy properties such as colour, magnitude and (relative) satellite magnitude, finding that the alignment is strongest for the reddest and brightest central and satellite galaxies. Our results confirm the theoretical picture and the role of the cosmic web in satellite accretion. Furthermore our results suggest that filaments identified on larger scales can be reflected in the positions of satellite galaxies that are quite close to their hosts.}
\end{abstract}

\begin{keywords}
methods: data analysis -- methods: statistical -- galaxies: statistics -- large-scale structure of Universe
\end{keywords}

\section{Introduction}

While the cosmic web \citep*{Joeveer:78,Bond:96} is interesting in and of itself, its characteristics affect the haloes and galaxies that inhabit it. Observations and simulations have shown that the spatial distribution of satellite galaxies is not random, but rather is aligned with the major axes of host galaxies. \citet{Tormen:97} was the first to suggest that galaxies fall into clusters/haloes anisotropically \citep*[see also][]{Knebe:04, Aubert:04}. According to current knowledge the satellites are preferentially (not randomly) accreted along filaments and retain a memory of the large scales from which they came \citep[e.g.][]{Knebe:04, Libeskind:11}.

The alignment between the distribution of satellite galaxies and large-scale structure (described e.g. by the velocity shear tensor) is affected mainly by two dynamical processes: a possible pre-adjustment of satellites in the filaments, which points radially toward the host galaxy/group; and the preferential accretion of satellites along those filaments. On smaller scales, the satellite distribution is also affected by the triaxial halo potential well.

Numerous studies have characterized different types of alignments between host galaxies and their satellites. These include the alignment of the spatial distribution of satellites with the orientation of the host halo/galaxy \citep[e.g.][]{Knebe:04, Zentner:05, Kang:07, Azzaro:07, Libeskind:07, Wang:08,Wang:10, Agustsson:10, Wang:14, Dong:14}; the alignment of the orientation of satellites/subhaloes with respect to the centre of the host \citep[e.g.][]{Faltenbacher:08, Schneider:13}; and the alignment between halo/group shape and the large-scale structure \citep[e.g.][]{Bailin:08,Paz:11,Zhang:13}. Again, accretion along the filaments and the impact of tidal fields have been invoked to explain the observed alignment signals. All these studies reveal a significant alignment between galaxies and their surrounding environments. The alignment is observed to be stronger for red galaxies/satellites and is almost absent between blue centrals and blue satellites.

Despite the numerous studies that analyse the correlations between satellites and host galaxies/haloes, only recently, the alignment between satellites and surrounding large-scale structure have been analysed. Using $N$-body simulations \citet{Libeskind:14} shows that the angular infall pattern of subhaloes is essentially driven by the shear tensor of the ambient velocity field. Dark matter (DM) subhaloes are found to be preferentially accreted along the direction of weakest collapse ($\mathbf{e}_3$ vector, the direction of filaments); a correlation is also found perpendicular to $\mathbf{e}_1$ (sheet normal). Similar conclusions is reached in \citet{Forero-Romero:14} where the Local Group like pairs are analysed in simulations. Also, the alignment of subhalo orbits is well correlated with the large-scale structure as defined by the velocity shear \citep{Libeskind:12, Libeskind:13}.

In observations the distribution of satellites and velocity shear are only analysed in nearby Universe. \citet*{Lee:14} show that the normal satellites and the dwarf satellites fall in to the Virgo cluster preferentially along the local filament and the local sheet, respectively. \citet{Libeskind:15} recently showed that planes of satellite galaxies observed around the Milky Way \citep[e.g.][]{Lynden-Bell:76, Lynden-Bell:82}, the Andromeda galaxy \citep{Ibata:13,Conn:13} and Centaurus~A \citep{Tully:15} are well correlated with the velocity shear field, reconstructed by surveys of cosmic flows. \citet{Cautun:15} showed that the planes of satellite galaxies are generally in very good agreement with predictions from \mbox{$\Lambda$CDM} simulations, albeit this was challenged by \citet{Ibata:14}. \citet{Pawlowski:12} argued that filamentary accretion cannot explain the orbits of the Milky Way satellites. Also, \citet{Forero-Romero:14} conclude that the Andromeda-Milky Way system is in contradiction to the accretion along the filaments. Thus the alignment between substructures and their large-scale structure is of great interest in the study of the local Universe.

Analysing the spatial distribution of satellites in larger samples is generally more challenging because only few satellites are detected per primary galaxy, and because the three-dimensional location of the satellites is uncertain. This measurement requires that primaries and their satellites be stacked to obtain statistically viable samples. Owing to the advent of large galaxy surveys, a statistically robust treatment of satellite galaxies has become possible \citep[e.g.][]{Guo:11, Guo:12, Wang:12, Wang:14a}.

In the current work we use the Sloan Digital Sky Survey (SDSS) to analyse the alignment between satellite distribution and the local filament/sheet orientation -- indirect tracers of the local velocity shear tensor. For filament detection we use the Bisous model, which is based on a marked point process \citep{Tempel:14c}. We check our results against semi-analytical models of galaxy formation run on merger trees extracted from numerical simulations. Such alignment exist in $N$-body simulations \citep{Forero-Romero:14,Libeskind:14} but is never searched in a large observational datasets, and hence never checked directly for consistency with numerical work.

The outline of the paper is following.  In Sect.~\ref{sec:data} we describe the data and methods and in Sect.~\ref{sec:res} we present our result. We conclude and discuss our results in Sect.~\ref{sec:con}. In Appendix~\ref{sec:bisous_vel} using high-resolution $N$-body simulation we show that the Bisous model filaments and sheets are aligned with velocity shear tensor.

\section{Data and Methods}
\label{sec:data}
  
  Throughout this paper we refer to isolated \textit{primary} galaxies (or just \textit{``primaries''}) as the central galaxies that host systems of fainter satellites and fulfil a set isolation criteria as in \citet*{Guo:15}. A satellites position vector is simply the vector pointing from the primary to a given satellite. We also refer to the alignment between the anisotropic distribution of satellites with filaments axes simply as the alignment, if not specified otherwise.

  \subsection{SDSS galaxies and filaments}
  
  Galaxies and their satellites from both the spectroscopic and photometric samples in the SDSS data release~8 \citep{York:00,Aihara:11} are used in this study. The catalogue of isolated primaries and their satellites that are in filaments are the same as that in \citet{Guo:15}. Primaries are selected from the spectroscopic sample, whereas satellite candidates can come from either the spectroscopic or photometric sample. We only consider isolated primaries in the sense that no other galaxy brighter than 0.5~mag less than the primary lies within a projected distance of $2R_\mathrm{inner}$ (where $R_\mathrm{inner}$ is described below) and is sufficiently close in redshift.

The limiting observed magnitude for satellite galaxies in the $r$~filter is 20.5. For a satellite with a spectroscopic redshift to be assigned to a host, the difference in the redshifts between the two must be less than 0.002. For satellites with only photometric redshifts, the difference between the photo-$z$ of the potential satellite and the primary's spectroscopic redshift of the host must be less than 2.5 times the photometric redshift error. All galaxies that are within a distance $R_\mathrm{inner/outer}$ from a primary are considered satellites if they satisfy the redshift and magnitude criteria mentioned above. Because of the poorly constrained distances provided when only photometric redshifts are available, given a candidate, it is not possible to determine if it is a genuine satellite or just a background/foreground galaxy (hereafter interloper) which is projected close to the primary. Statistically, genuine satellites tend to reside close to their primaries. As in \citet{Guo:15}, we examine galaxies in the inner area (within a projected distance $R_\mathrm{inner}$ of a primary galaxy, hereafter ``inner'' galaxies) and outer area (the annulus bounded by the projected distance $R_\mathrm{inner}$ and $R_\mathrm{outer}$ of the same primary, hereafter ``outer'' galaxies). The values of $R_\mathrm{inner}$ and $R_\mathrm{outer}$ are the same as in \citet{Guo:15}, which depend on the primary's magnitude. We divided the primary sample into three $r$-band magnitude bins, each one magnitude wide and centred on $M_r$ = $-21.0$, $-22.0$, and $-23.0$. The values of $(R_\mathrm{inner}, R_\mathrm{outer})$ are $(0.3, 0.6)$, $(0.4, 0.8)$, and $(0.55, 0.9)$~Mpc, respectively. More details about primary and satellite galaxy selection are given in \citet{Guo:11,Guo:12}.

  As shown in \citet{Guo:13, Guo:15}, the number density of genuine satellites in the outer areas is usually lower than that in the inner area, since interloper galaxies dominate the outer areas.  Also, interlopers  are distributed randomly with respect to the filaments. There should be no correlation between the angular distribution of interlopers and the local filaments axis. Therefore the alignment signals for outer galaxies  must be diluted by these  isotropically distributed interlopers. The details of measurement of the alignment signals and handling of interloper galaxies are described in Sect.~\ref{sec:method}.
    
  The catalogue of filaments is built by applying the Bisous process to the distribution of galaxies as described in Sect.~\ref{sec:bisous}. The method and parameters are exactly the same as in \citet{Tempel:14c}. The assumed scale (radius) for the extracted filaments is 0.71~Mpc. Because the survey is flux-limited, the sample is very diluted farther away. Hence, we are only able to detect filaments in this scale up to the redshift 0.15 ($\approx 640$~Mpc). We refer to \citet{Tempel:14c} for the details of our filament finder. In addition to the filament direction ($\mathbf{e}_3$ vector), we also estimate the orientation of sheets ($\mathbf{e}_1$ and $\mathbf{e}_2$ vectors) in the location of filaments as described in Sect.~\ref{sec:bisous_b1}.
  
  In our analysis, we use only primary galaxies that are located in a filament, i.e. the primary distance from filament axis is less than 0.71~Mpc and the distance of the galaxy from the end point of filament (if the galaxy is outside a filamentary cylinder) is less than 0.14~Mpc.

  \subsection{Semi-analytical galaxies and filaments}
  
  The model galaxy sample was created using the publicly available Millennium Simulation \citep[MS,][]{Springel:05} to define the mass distribution and a semi-analytic model to place the galaxies within this density field. The simulation covers a large volume with a box length of 714~{Mpc}. The mass resolution used is $\sim 10^9~M_{\odot}$ per particle.  Galaxies, whose properties are computed semi-analytically populate the DM haloes according the galaxy formation model {\sc galform} \citep{Bower:06}. It includes re-ionisation at high redshift, feedback from supernovae and stellar winds (in order to prevent the overproduction of the low luminosity galaxies), and outflows from active galactic nuclei (AGN) to suppress the formation of most luminous galaxies. The model was tuned to reproduce the observed $K$-band luminosity at redshift $z \sim 0$. The low absolute magnitude limit of our model galaxy sample is about $-16$ in $r$ band: this sets the magnitude limit for potential satellite galaxies. Fig.~5 of \citet{Guo:13} demonstrated that such resolution is good enough for a reliable comparison between the simulation's satellite population and those in SDSS data.
  
	In the semi-analytical model, isolated primaries and their satellites are selected as described in sect.~3 of \citet{Guo:13}, which is slightly different to the procedures applied to the SDSS light-cone survey data but with same physical motivation. The slight difference arises because the semi-analytical galaxies are in real-space while SDSS galaxies are in redshift-space. For semi-analytical galaxies in magnitude bins, $-21.0$, $-22.0$, and $-23.0$, we define the term ``inner'' to mean within the sphere of the radius, 0.3, 0.4 and 0.55~Mpc, respectively. The outer galaxies are those found in between $R_\mathrm{inner}$ and $R_\mathrm{outer}$, where $R_\mathrm{outer}$ is 0.6, 0.8, and 0.9~Mpc. The values of $R_\mathrm{inner}$ and $R_\mathrm{outer}$ are the same as for SDSS galaxies, except they are measured in real space rather than in projection. The remaining isolation criteria for the semi-analytical galaxies are also the same as for SDSS galaxies. Since the inner and outer galaxies are selected in real space, there are no background/foreground interlopers among the satellite galaxies in this case. The simulation is thus used as a tool to identify the uncontaminated ``true'' signal.
  
  We have shown in \citet{Guo:13} that the estimated satellite luminosity functions and projected number density profiles from light-cone data (for SDSS sample, with background subtraction) is not biased from those estimated in real space (for the model galaxy sample, without background subtraction). The estimated satellite luminosity functions and projected number densities are very similar. In this paper, we will directly compare the measured alignment signals from light-cone data (for the SDSS sample) with the projected 3D position (for the semi-analytical galaxy sample) as well. It is convenient to explore the dependence of the alignment signal on various properties for the simulation data, since there are (by constructions) no interlopers in the inner or outer areas. Because of the interlopers, the alignment signal for SDSS galaxies is necessarily weaker in most cases \citep[e.g. see][]{Agustsson:10} compared with that from semi-analytical galaxies. That said, we do not explicitly attempt to recover the observational signal from the semi-analytical sample. Instead, we attempt to understand the underlying dependencies of the alignment on galaxy properties. Therefore, in what follows we compare our results from the SDSS light-cone survey directly with those from semi-analytical galaxies, having in mind that we do not expect these two to be identical.
  
  The same filament finder  is applied to both SDSS and semi-analytical galaxies to get the filaments catalogue. For the Bisous model, we use only galaxies brighter than $-19.0$~mag in $r$ band. In principle we could also use fainter galaxies to extract the filamentary network, however, we wish to avoid using satellites when extracting filaments. In this way, the alignment between satellite distribution and Bisous filaments and sheets is not built into the method (and hence trivial). 
  
  For filament extraction, we use the same parameters for the Bisous model as we used for SDSS (see previous section). Primaries are classified as ``in-a-filament'' or ``not-in-a-filament'' in the same way for both samples.

\subsection{Bisous model for filament extraction}
\label{sec:bisous}

We apply an object/marked point process with interactions \citep*[the Bisous process;][]{Stoica:05} to trace the filamentary network in the distribution of galaxies. This algorithm provides a quantitative classification which agrees with the visual impression of the cosmic web and is based on a robust and well-defined mathematical scheme. A thorough explanation of the method can be found in \citet*{Stoica:07,Stoica:10} and \citet{Tempel:14c}; for convenience, a brief description is given below.

The marked point process we use for filament detection do not model galaxies/haloes, but models the structure outlined by galaxy/halo positions. This model approximates the filamentary network by a random configuration of small segments. We assume that locally galaxies may be grouped together inside a rather small cylinder, and such cylinders may combine to form a filament if neighbouring cylinders are aligned in similar directions.

The solution provided by our model is stochastic. Therefore, we find some variation in the detected patterns for different Markov chain Monte Carlo (MCMC) runs of the model. The main advantage of using such a stochastic approach is the ability to give simultaneous morphological and statistical characterization of the filamentary pattern.

In practice, after fixing the approximate scale of the filaments, the algorithm returns the filament detection probability field together with the filament orientation field. Based on these data, filament spines are extracted and a filament catalogue is built. Every filament in this catalogue is represented as a spine: a set of points that define the axis of the filament.

The spine detection we use is based on two ideas. First, filament spines are located at the highest density regions outlined by the filament probability maps. Second, in these regions of high probability for the filamentary network, the spines are oriented along the orientation field of the filamentary network. See \citet{Tempel:14c} for more details of the procedure.

\subsubsection{Complementing the Bisous filaments with sheet orientation}
\label{sec:bisous_b1}

As already mentioned, the Bisous model is probabilistic and gives the filament detection probability field (called the ``visit map''). At the location of filament spines, the cross section of the visit map can be analysed. Assuming that filaments are located in sheets, the distribution of haloes/galaxies in filaments is elongated in the plane of the sheet. The real space elongation of haloes/galaxies will also affect the filament detection probability field. Hence, the cross section of visit map in the location of filaments should be elongated in the plane of the sheet. Consequently, the probabilistic nature of Bisous model can be exploited to define the sheet normals in the location of filaments. This idea was already employed in \citet{Tempel:13}, where the orientation of galaxies with respect to the Bisous filaments and sheets were analysed.

\begin{figure}
	\centering
	\includegraphics[width=84mm]{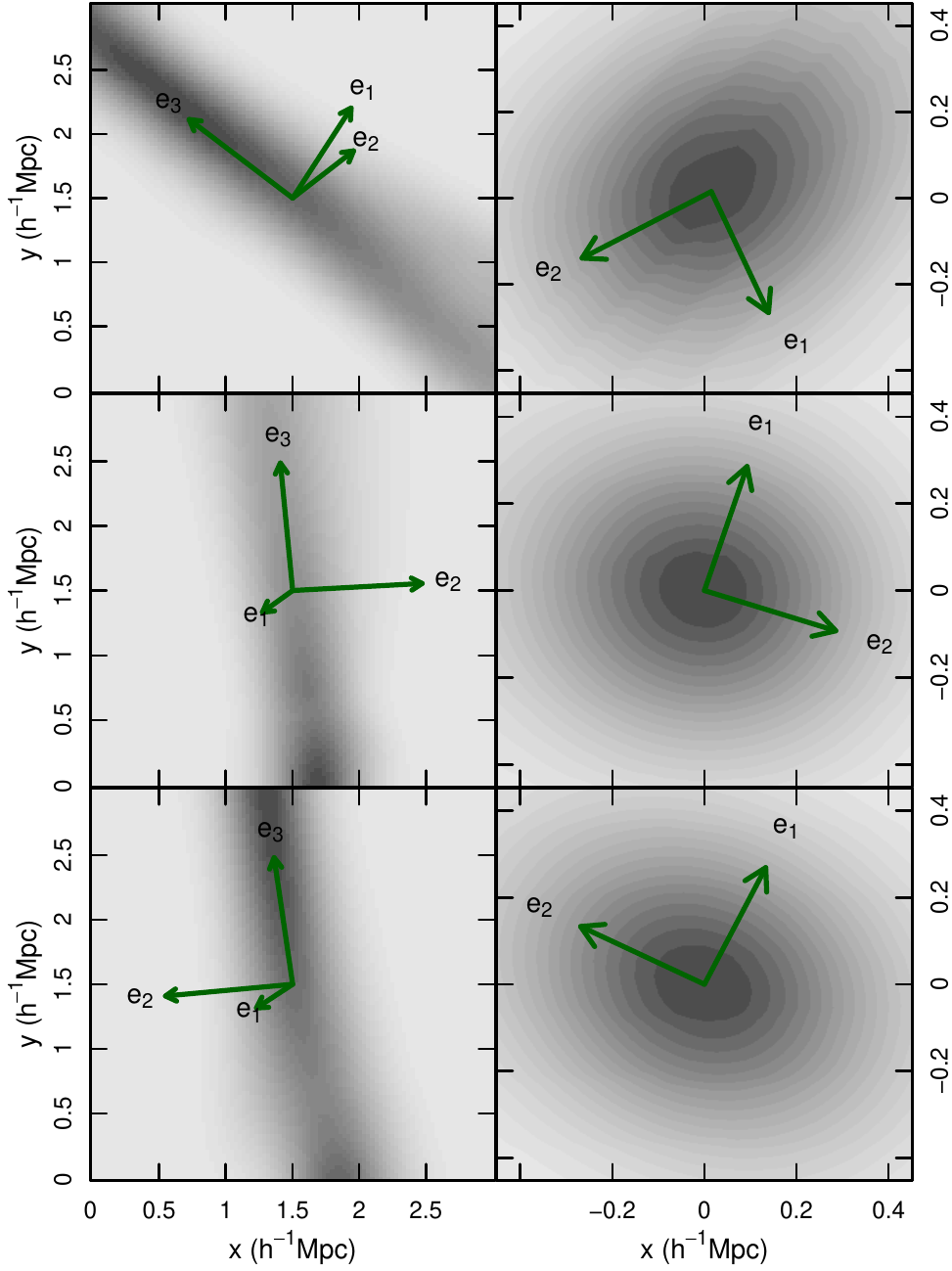}
    \caption{Illustration of Bisous model filaments and the large-scale structure axes defined by the Bisous model. Three examples of filaments and their cross-sections are shown. Left panels show the filaments defined by the visit map. The filament direction is indicated with $\mathbf{e}_3$ vector, whereas $\mathbf{e}_1$ and $\mathbf{e}_2$ vectors are perpendicular to it. Right panels show the cross-section of those filaments and the axes $\mathbf{e}_1$ and $\mathbf{e}_2$. The axis $\mathbf{e}_1$ is defined as the short semi-axis of the cross-section in the visit map. The axis $\mathbf{e}_2$ is perpendicular to $\mathbf{e}_1$ and $\mathbf{e}_3$. The axis $\mathbf{e}_1$ defines the sheet normal, where the filament is located.}
     \label{fig:Bisous_example}
\end{figure}

In Fig.~\ref{fig:Bisous_example} we show the examples of Bisous filaments defined using the visit map. In the right column the cross section of a filament is shown. The elongation of the visit map is modest, but this is expected since we use a cylindrical shapes to model the filamentary network and most importantly, initially the Bisous model was designed to detect filaments not sheets.

We designate the orientation of Bisous filaments as $\mathbf{e}_3$, whereas $\mathbf{e}_2$ and $\mathbf{e}_1$ are perpendicular to filaments and each other, and the latter defines the normal of the sheet defined using the Bisous model (see Fig.~\ref{fig:Bisous_example}).

As we show in \citet{Tempel:14b} and Appendix~\ref{sec:bisous_vel}, the $\mathbf{e}_i$'s defined in the Bisous model are very well aligned with the eigenvectors of the velocity shear field, hence, the Bisous model can be used as an indirect tracer of the velocity shear.

  \subsection{Measuring the alignment signal}
  \label{sec:method}
  
In order to examine (and quantify) if satellites are anisotropically distributed with respect to filaments, the probability function $P(\theta)=N(\theta)/\langle N(\theta)\rangle$ (for both SDSS and semi-analytical galaxies) is measured. The alignment signal is calculated using the kernel density estimation\footnote{In Appendix~A of \citet{Tempel:14d} we show that the kernel density estimation is better than a simple histogram and is more representative of the underlying probability distribution.}. Here $\theta$ is the angle between the filament axis and the satellite's position vector (relative to its host). For SDSS galaxies, only the component of $\theta$ projected in the sky plane can be measured. For the semi-analytical sample, we project the relative satellite position onto an arbitrary plane (namely, one defined by two of the simulation's box axes) and measure the angle $\theta$ between the projected off-centre vectors and filament axes. This is then compared with the measurements from the SDSS sample. $N(\theta)$ and $N(\cos\theta)$ are the number of the central-satellites pairs that subtend an angle $\theta$ from the local filamentary structure's main defining axis. $\theta = 0\degr$ or $\cos\theta = 1$ implies that the satellites are preferably distributed along the filament's axis (hereafter, referred to as ``alignment''), while $\theta = 90\degr$ or $\cos\theta = 0$ implies that satellites are preferably distributed perpendicular to this  filamentary axes (hereafter termed ``anti-alignment'').

To calculate the alignment signal, we use all galaxies in the inner or outer regions. For semi-analytical galaxies, all galaxies in inner or outer regions are close to the central galaxy. For SDSS the inner and outer galaxy samples also include interloper galaxies that actually do not belong to the central galaxy. For the calculation of alignment signal, we do not try to remove interlopers or subtract any background/foreground. Instead, we show the alignment signals for both inner and outer galaxies (using all galaxies in inner and outer regions)\footnote{Note that interlopers mostly affect the strength of the alignment signal. In general, if the strength of the alignment signal is important, the background/foreground should be statistically taken into account \citep[e.g. see][]{Wang:12,Cautun:15}.}.

  To estimate the significance of the alignment, we need to compare the measured probability function to the null-hypothesis of random (isotropic) angular satellite distribution. However, due to selection effects, it is not granted that a random angular satellite distribution \citep*[nor the distribution of filaments axes with respect to the line-of-sight,][]{Tempel:13a} results in a uniform probability distribution. Therefore we use a Monte Carlo (MC) approximation similar to the method used in \citet{Tempel:13a} to estimate the case that the angular distribution of the satellites is random, regardless of filament axes and the confidence intervals for this estimate. This approach will take simultaneously into account the biases in the filament detection and angular distribution of satellites.

  In order to do so, we generate 400 randomised samples (for the subsamples of relatively small volume, such as blue primaries, we generate 600 randomised samples) in which satellite position vectors with respect to their primaries are kept fixed, but the filament axes of the primary are shuffled and re-assign to each primary. The median and its 5\% and 95\% quantiles are then calculated by using these random samples. These confidence regions together with the Kolmogorov-Smirnov test (KS test) probabilities that the angle distributions are drawn from a uniform distribution are shown in the figures.

\begin{figure*}
	\includegraphics[width=176mm]{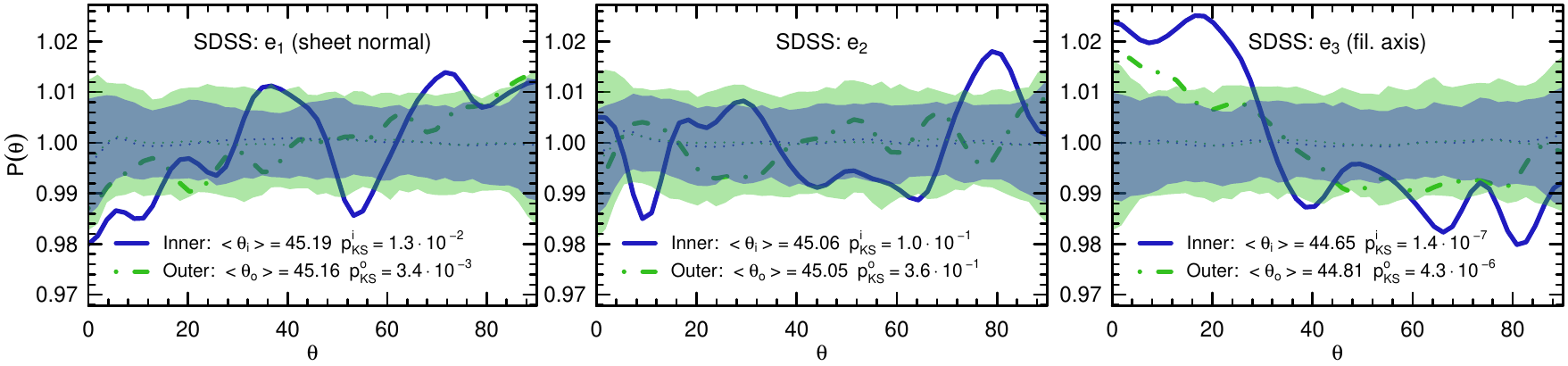}
    \caption{The alignment of satellites in SDSS as a function of the angle $\theta$ (measured in the plane perpendicular to the line of sight) between the satellite position vector (relative to the primary galaxy) and the orientation of structures ($\mathbf{e}_i$) as defined in the Bisous model. The vector $\mathbf{e}_3$ (right panel) designates the filament orientation, while $\mathbf{e}_1$ (left panel) gives the normal of the sheet where the filament is located. The alignment is shown for inner (close to the primary) and outer (farther away from the primary) satellites. The definition for inner and outer satellites is given in the text. The filled regions show the 5th and 95th percentile spread for a randomised distribution. On each panel, we also give the average angle and the Kolmogorov-Smirnov (KS) test probability that the angle distribution is drawn from a randomised distribution. Satellite galaxies in the SDSS are statistically well aligned with the filamentary structure $\mathbf{e}_3$.}
     \label{fig:sat_sdss}
\end{figure*}

\begin{figure*}
	\includegraphics[width=176mm]{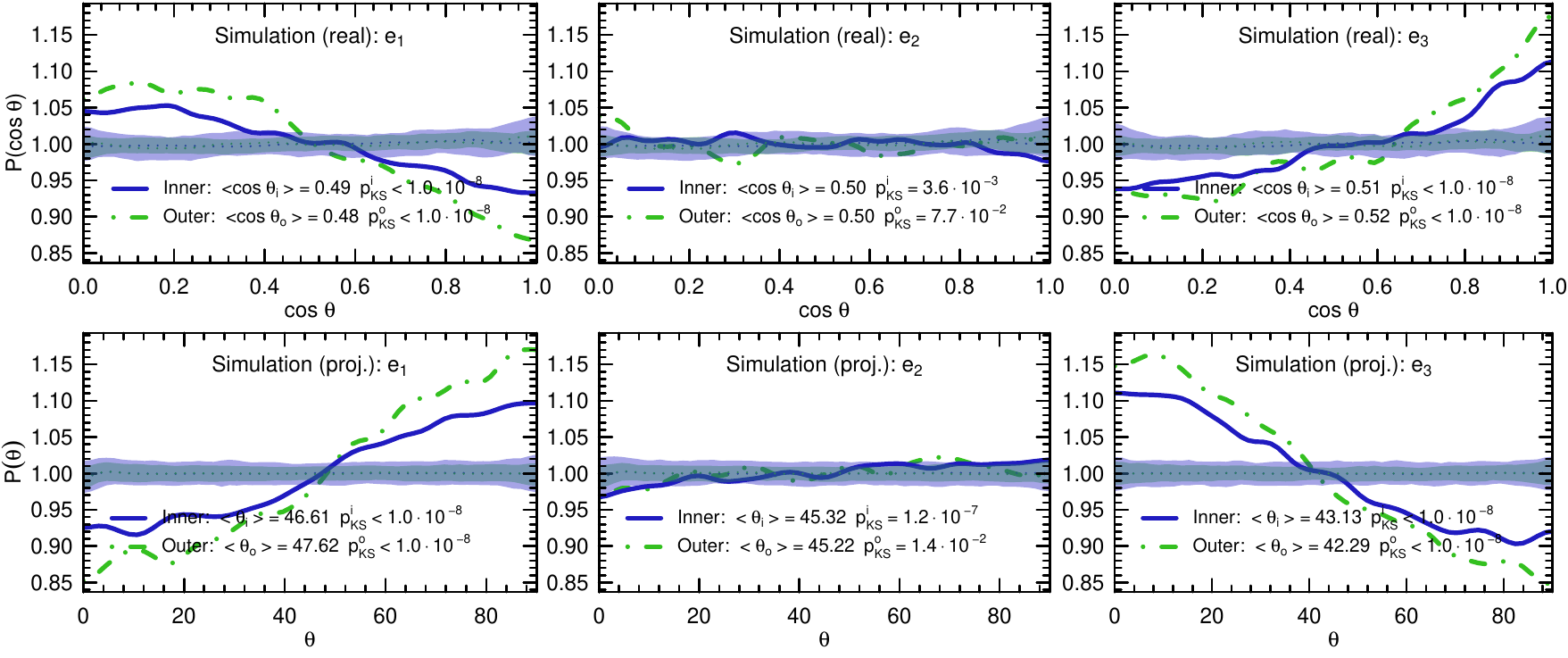}
    \caption{The real (upper row) and projected (lower row) space alignment between anisotropic distribution of satellites and $\mathbf{e}_i$ vectors in Millennium simulation. The definition of $\mathbf{e}_{i}$ is given in Sect.~\ref{sec:bisous_b1}; $\mathbf{e}_3$ gives the filament orientation and $\mathbf{e}_1$ defines the normal of the sheet where the filament is located. The designation of lines and labels are the same as in Fig.~\ref{fig:sat_sdss}.}
     \label{fig:sat_ms}
\end{figure*}
  
\section{Results}
\label{sec:res}

\subsection{Satellite alignment in SDSS}
\label{sec:sdss}

  In Fig.~\ref{fig:sat_sdss}, we show the probability distribution $P(\theta)$ for the angle (in the sky plane perpendicular to the line of sight) between satellite position vectors and the filament/sheet axes ($\mathbf{e}_i$) for SDSS galaxies. Although the $R_\mathrm{inner}$ varies with the magnitude of primaries, for the angle $\theta$ we can measure the mean excess probability distribution $P(\theta)$ for all primaries (including all galaxies in three magnitude bins $M_{\mathrm{p}}^r=-21, -22, -23$) by simply stacking them together. We will explore the possible dependence of the excess probability on the properties of galaxies later. In each panel of Fig.~\ref{fig:sat_sdss}, we show the average angle $\langle\theta_\mathrm{i/o}\rangle$ and the Kolmogorov-Smirnov (KS) test probability $p_\mathrm{KS}^\mathrm{i/o}$ that the sample is drawn from the randomised distribution for inner/outer galaxies.

Fig.~\ref{fig:sat_sdss} shows that the satellites around primaries are preferentially distributed along the filament axis ($\mathbf{e}_3$ vector) and have anti-alignment with the sheet normal ($\mathbf{e}_1$ vector) where the filament is located. This results is in accordance with the predictions from pure $N$-body simulation \citep{Libeskind:14}. Similarly in simulations, the alignment between satellite distribution and $\mathbf{e}_1$, $\mathbf{e}_3$ vector is statistically significant (see the KS test values), while the alignment between satellites and $\mathbf{e}_2$ vector is very weak or absent.

In order to compare our results with simulations, in the next section we calculate the same alignments for the galaxies in the Millennium simulation.

\subsection{Satellite alignment in Millennium simulation}
\label{sec:mill}

\begin{figure*}
	\includegraphics[width=176mm]{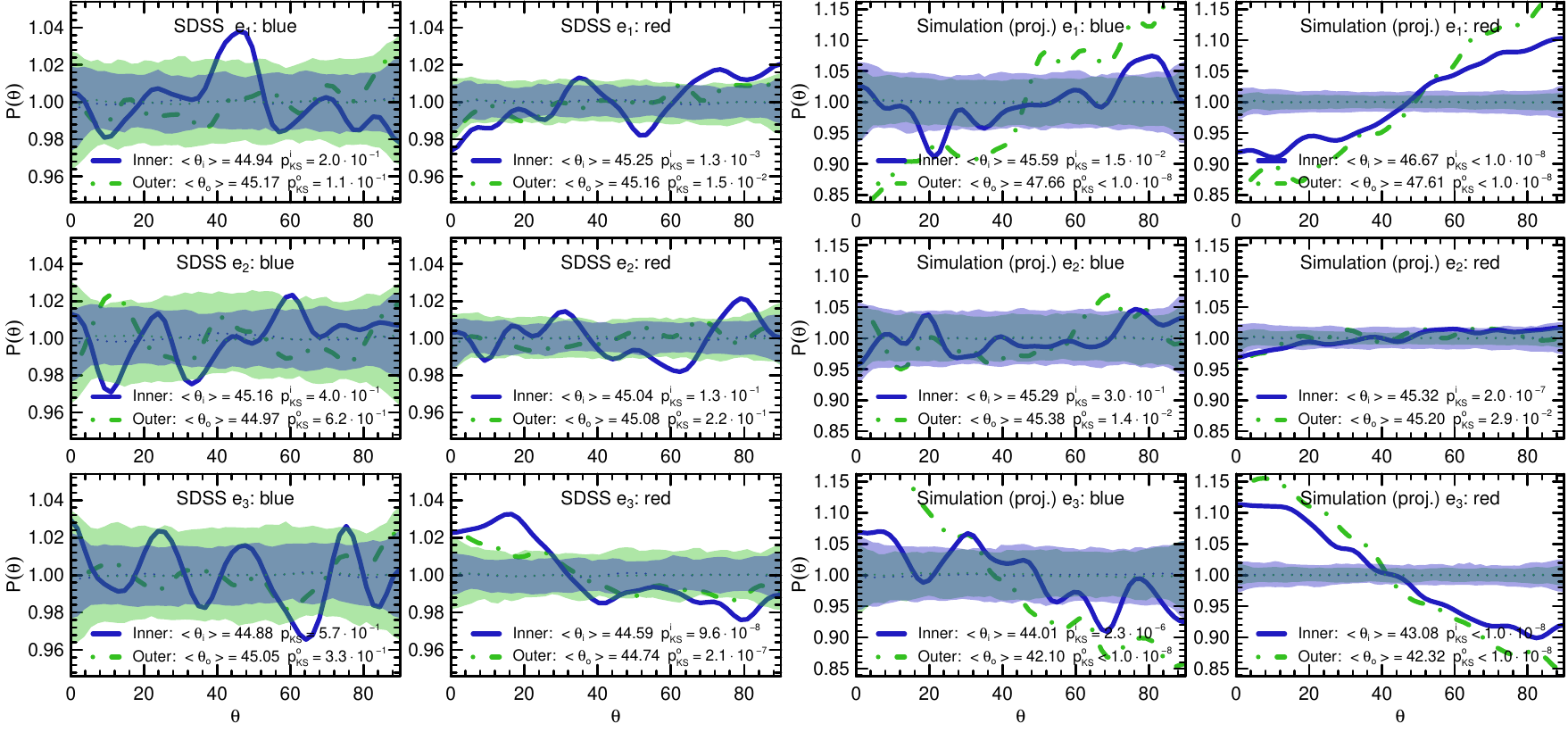}
    \caption{The alignment between anisotropic distribution of satellites and $\mathbf{e}_i$ vectors in SDSS (two left-hand columns) and in Millennium simulation (two right-hand columns). The alignment is shown for red and blue primaries. The designation of lines and labels are the same as in Fig.~\ref{fig:sat_sdss}.}
     \label{fig:sat_colour}
\end{figure*}

In this section we examine satellite positions with respect to the $\mathbf{e}_i$ vectors as defined by Bisous model using the Millennium simulation. Fig.~\ref{fig:sat_ms} shows the alignment between satellites and $\mathbf{e}_i$ vectors in real space (upper row) and in projection (lower row). In both cases a statistically significant alignment is seen, and in qualitative agreement with the alignment measured for the observed SDSS galaxies. The strength of the alignment between satellite positions and the $\mathbf{e}_i$ vectors is  (obviously) weakened when computed in projection; it is however not erased.

For both SDSS and semi-analytical primaries, the alignment results show similar trends that the satellites around them preferentially distribute along the filaments axes and have anti-alignment with the sheet normal. For SDSS galaxies, this alignment is stronger for inner galaxies than for outer galaxies (see Fig.~\ref{fig:sat_sdss}), which is opposite for semi-analytical galaxies (Fig.~\ref{fig:sat_ms}). The signals of alignment for outer galaxies in SDSS sample could be diluted because most galaxies in the outer areas are interlopers whose distribution is isotropic with respect to the filament axis. However, for the semi-analytical galaxy sample, the outer galaxies are actually relevant to the primaries -- they are not far from the primaries in distance. Furthermore, the distribution of these outer galaxies may be influenced directly by the filaments. It is possible that these outer galaxies are themselves also preferentially distributed along the filament axis and are even more strongly aligned with it. This is because the distribution of outer galaxies is less affected by the primary's DM halo (compared with inner galaxies that may well be orbiting within the host's DM halo). After galaxies are accreted onto the main DM halo, the ``memory'' of the alignment with the filament axis may be forgotten or weakened \citep[e.g. see][]{VeraCiro:11}. This interpretation is in accordance with the alignment of galaxy pairs in filaments. Namely, \citet{Tempel:15} shows that the alignment of loose galaxy pairs with filaments is much stronger than that for close galaxy pairs.

Recall that the strength of the alignment measured from the SDSS and Millennium simulation sample is expected to be different (see Sect.~\ref{sec:method}). The strength of alignment depends on several factors, (examined below) including the number of background/foreground galaxies in the inner or outer areas, the properties of primaries and satellites used for the measurement.

  \begin{figure}
  	\includegraphics[width=84mm]{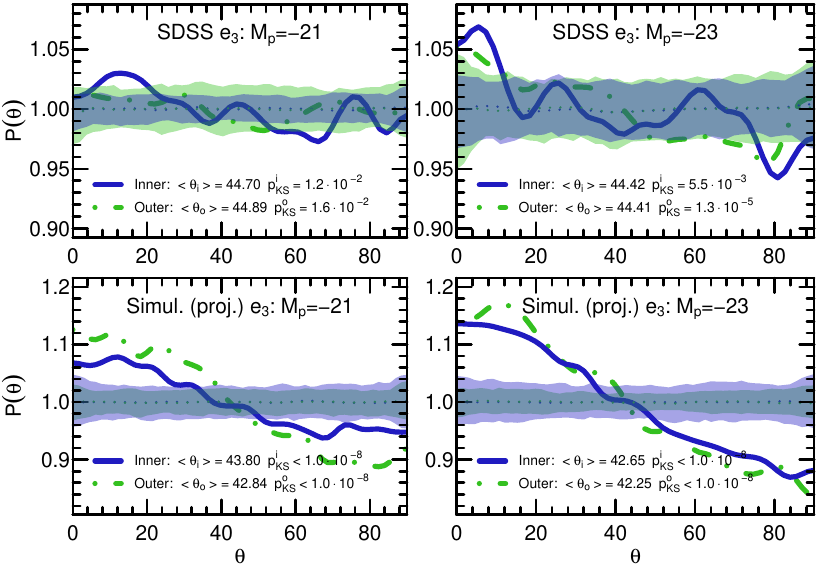}
      \caption{The alignment between anisotropic distribution of satellites and filament orientation ($\mathbf{e}_3$ vector) in SDSS (upper row) and in Millennium simulation (lower row). The alignment is shown for bright ($M_\mathrm{p}=-23$) and faint ($M_\mathrm{p}=-21$) primaries. The designation of lines and labels are the same as in Fig.~\ref{fig:sat_sdss}.}
       \label{fig:test5}
  \end{figure}

  \begin{figure}
  	\includegraphics[width=84mm]{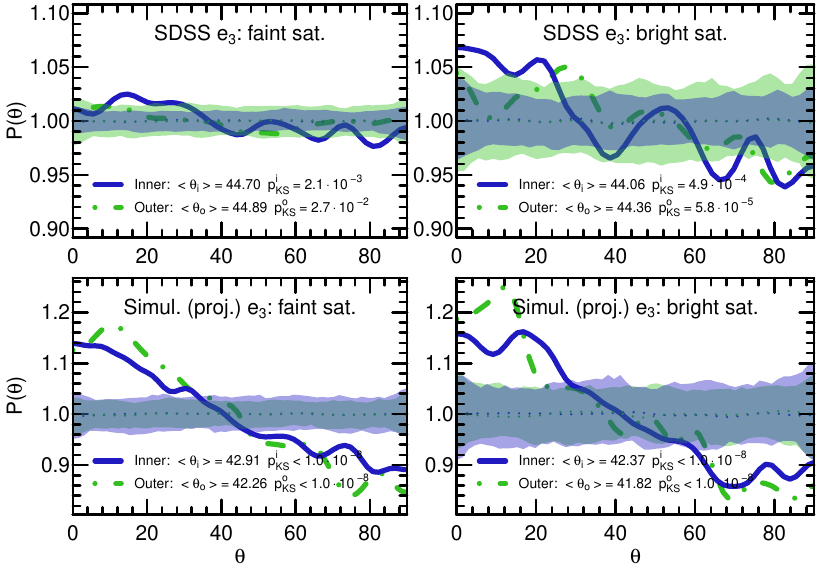}
      \caption{The alignment between anisotropic distribution of satellites and filament orientation ($\mathbf{e}_3$ vector) in SDSS (upper row) and in Millennium simulation (lower row). The alignment is shown for bright and faint (relative to the luminosity of primary) satellites. The designation of lines and labels are the same as in Fig.~\ref{fig:sat_sdss}.}
       \label{fig:test4}
  \end{figure}

\subsection{Dependence on galaxy properties}

  In this section, we split our primaries into different subsamples. We then explore the possible dependence of the alignments on the properties of primaries or satellites.

\subsubsection{Colour dependence of alignment signal}

  Firstly, we split the primaries into subsamples by ${}^{0.0}(g-r)$ colour of the primaries. We use two different lines, ${}^{0.0}(g-r) = 0.15-0.024~M_r$ and ${}^{0.0}(g-r) = -0.28-0.04~M_r$ to divide the primaries into red and blue subsamples for SDSS and semi-analytical galaxies, respectively, due to the distribution of SDSS galaxies in the colour-magnitude diagram is shifted along the $g - r$ axis compared to that of the semi-analytical galaxies \citep[refer to][for the choice of the lines]{Guo:13}. Fig.~\ref{fig:sat_colour} shows the measured alignment for red and blue primaries respectively. For the red primaries in both samples, the strength of alignment of inner and outer galaxies are both very pronounced. The KS test probability for distribution of inner galaxies around SDSS red primaries is as low as $\sim 10^{-7}$. However, for SDSS blue primaries, the result of angular distribution of inner galaxies and outer galaxies with respect to filament axes is statistically isotropic (Fig.~\ref{fig:sat_colour}, bottom left panel), which is different to the results from the simulation (Fig.~\ref{fig:sat_colour}, panel on the bottom row, third column). For model blue primaries, the inner galaxies around them show a weak preference of distributing along the filament axes, while a strong preference of distributing along the filament axes for outer galaxies. This suggest that the alignment for outer galaxies might be weakened when they fall into the main halo (and became inner galaxies). In SDSS the alignment is not observed, probably because of interloper galaxies in the sky projection.
  
  In Fig.~\ref{fig:sat_colour} we show the alignment for all $\mathbf{e}_i$ vectors and the dependence on colour is the same for all of them. Since the dependence is strongest for $\mathbf{e}_3$ vector, we continue by presenting only this alignment. The alignment for $\mathbf{e}_1$ is similar, but weaker and omitted in the interest of clarity.

\subsubsection{Dependence on primary magnitude}

We split the primaries into subsamples by magnitude and explore the dependence of the alignment on this quantity. Fig.~\ref{fig:test5} shows the resulting alignments for these subsamples. For both SDSS and model galaxy subsamples, the strength of alignment slightly depends on the luminosity of primaries; although for SDSS galaxies, different number of interlopers could also cause the variation in the strength. The $\langle\theta_\mathrm{i}\rangle$ for inner galaxies decrease from 44.70 to 44.42 from primaries of low luminosity to high luminosity. The $\langle\theta_\mathrm{i}\rangle$ for primaries in magnitude bin $M_{\mathrm{p}}^r=-22$ is 44.70, which is roughly the same as that for magnitude bin $M_{\mathrm{p}}^r=-21$. However the change of overall shape of distribution $P(\theta$) with the primary luminosity is consistent with our previous statement. Moreover, the results from semi-analytical galaxies, which is not affected by interlopers, also show similar dependence: the strength of the alignment will increase with the luminosity of primaries. The luminosities of primaries are loosely correlated with their masses. Therefore, we could also say that the strength of alignment depends on the mass of the primary galaxy \citep[analogous to the trend found for DM halo shapes in][]{Libeskind:13}.

\subsubsection{Dependence on satellite magnitude}

  Here, we split the satellites around the all primaries into subsamples by the luminosity of satellites, bright satellites $M_\mathrm{sat} < M_\mathrm{p} + 2$, middle luminous satellites $M_\mathrm{p} + 2 < M_\mathrm{sat} < M_\mathrm{p} + 3$, and faint satellites $M_\mathrm{p} + 3 <M_\mathrm{sat}< M_\mathrm{p} + 4$. Then we explore how the alignment depends on the relative satellite-host luminosity. Fig.~\ref{fig:test4} shows the alignment for satellites in bright and faint subsamples. Again, both results from SDSS and model samples imply that the strength of alignment slightly depends on the luminosity of satellites. The $\langle\theta_\mathrm{i}\rangle$ for three SDSS subsamples (model subsamples) are 44.06, 44.38 and 44.70 (42.37, 42.55, 42.91), which is anti-correlated with satellites luminosity. The trend for $\langle\theta_\mathrm{o}\rangle$ is the same in both cases. This shows that the alignments for bright satellites are stronger. The results also indicate that the measured alignment with respect to $\mathbf{e}_i$ is not affected by the faint magnitude limit in the SDSS and in simulations: the alignment is qualitatively the same regardless of the luminosity of satellites.

  \subsection{Intrinsics of the alignment signal}

At this point in our study it is relevant to highlight that numerous observational studies \citep[since the seminal work of][]{Holmberg:69} have examined the angular distribution of satellites around primaries \citep[][although not all of these studies are in full agreement]{Zaritsky:97, Brainerd:05, Agustsson:06}. Perhaps the most relevant work in this context is \citet{Yang:06} who found that satellites in the SDSS tend to be found along the main axis of their host galaxies. We find similar alignment (see Appendix~\ref{sec:ps}). On the other hand, the major axis of galaxies was also found to be correlated with the axes of filaments \citep[e.g.][]{Tempel:13a}, which is again corroborated here in Appendix~\ref{sec:pf}. Therefore we wish to explore if the alignment between the anisotropic distribution of satellites and filament axes shown above found here is intrinsic or just the result of the collective effect of two former alignments that are already well known. Hereafter, the known alignments are referred as ``established'' alignments.

\begin{figure}
	  \centering
	  \includegraphics[width=80mm]{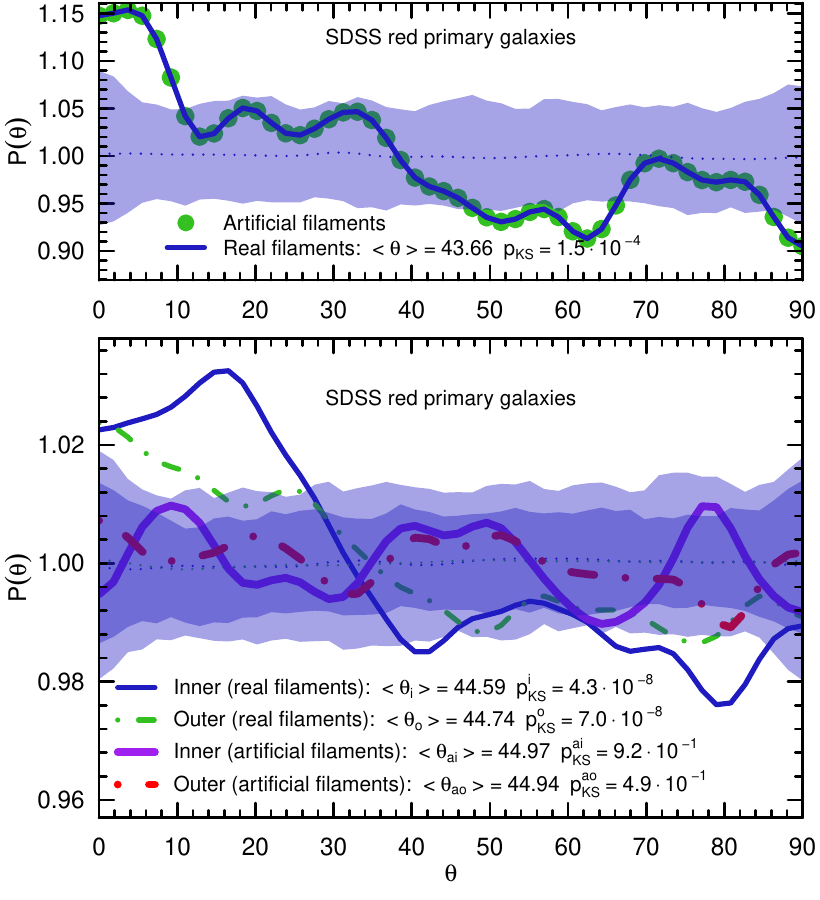}
    \caption{Test of the intrinsics of the alignment between the filament axes and anisotropic distribution of satellites. \textit{Top panel:} the alignment between real/artificial filament axes and the major axis of red primaries. \textit{Bottom panel:} the alignments between anisotropic distribution of satellites with artificial filament axes (thick red and purple lines) and with real filament axes (thin blue and green lines).}
    \label{fig:fil_intrinsic}
  \end{figure}

In order to check this we collect all the angles between the filament axes and primary galaxy major axes in the plane of the sky. Then for each primary, we randomly select a new angle from this distribution. We thus generate an artificial filament axes for each primary according to this new picked angle\footnote{An alternative possibility is to randomise the primary galaxy major axes with respect to filaments, while fixing the filament orientation and satellite distribution. In this case, the alignment between artificial major axes and satellite positions should be analysed. Statistically speaking, these two approaches are identical in the ideal case that all satellite are aligned with filament axes (since in both cases same angles are randomised).}. The resulting alignment signal between these artificial filament axes and the primary major axes is by construction statistically identical to that with the real filaments as shown in the top panel of Fig.~\ref{fig:fil_intrinsic}. A crucial aspect of this test is that it leaves the alignment between the satellite position and galaxy major axis, undistributed. Therefore any direct correlation between the filaments and the anisotropic distribution of satellites is randomised, and the remaining correlation is due to the combined effect of two established alignments. If the previous alignments between real filament axes and satellite positions is intrinsic, the alignment signal will be much weaker when artificial filaments axes are considered, since the direct correlation is broken. If the satellite position -- filament axis alignment is just the consequence of the two established alignments mentioned earlier, a similar alignment signal should also exist when the artificial filaments are considered, since these two established alignments are preserved and statistically same as for real filaments.

Suppose satellite positions are locked to the major axis of their host galaxy. In this case randomising the angle between the filament and the host major axis will have the effect of randomising the angle between the filament and the satellite distribution. In this case however, the distribution of angles between the satellite galaxy position and the filament will not change since by construction the distribution of angles between the galaxy and filament has not changed. Now if on the other hand, satellite galaxies are locked to the filaments, then randomising the angle between filament and host major axis will have (1)~no  effect on the satellite galaxy position-host major axis alignment (this has not been touched) and (2)~no effect on the galaxy major axis-filament alignment (by construction). The effect will only be to erase the direct alignment between the satellites and the filament. The remaining alignment is the collective effect of two established alignments.

  \begin{figure}
  	\includegraphics[width=84mm]{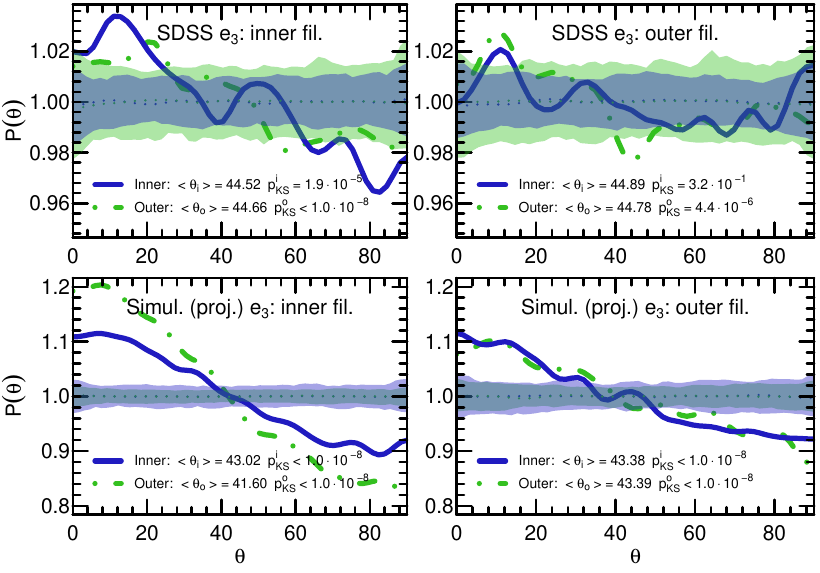}
      \caption{The alignment between anisotropic distribution of satellites and filament orientation ($\mathbf{e}_3$ vector) in SDSS (upper row) and in Millennium simulation (lower row). The alignment is shown for primary galaxies that are in inner part of filaments or in outer part of filaments. The designation of lines and labels are the same as in Fig.~\ref{fig:sat_sdss}.}
       \label{fig:fildist}
  \end{figure}

  The result of alignment between the artificial filament axes and the major axis of red primaries is shown in the bottom panel of Fig.~\ref{fig:fil_intrinsic}. It shows there is statistically no alignment (or the alignment is very weak) between the artificial filament axes and anisotropic distribution of satellites, which implies that the well established alignments are not sufficient to explain the measured alignment between the anisotropic distribution of satellites and filament axes. In other words: \emph{the alignment of satellites with the filamentary structure is intrinsic.}

 As one final check, we split our primaries according to their distance to the filament axis and examine the satellite alignment, since \citet{Tempel:13} also found the dependence of this property on major axis-filament alignment: the alignment is stronger in outer parts of filaments. The results are shown in Fig.~\ref{fig:fildist}. For galaxies further away from the filament axis, the satellite-filament alignment is weaker (even for red primaries). This is another indicator that the alignment we found is not trivial results of combined effects, since otherwise the alignment should be stronger in outer parts of filaments (since the alignment between galaxy major axis and filaments is stronger in outer parts).

\section{Discussion and conclusions}
\label{sec:con}

In the present paper, we extended the probabilistic model for filament detection \citep[called Bisous model;][]{Tempel:14c} and showed that the stochastic nature of Bisous model can be used to trace the orientation of large-scale structure in the location of galactic filaments. We analysed the alignment between satellite distribution around host galaxies and cosmic filaments in the Millennium simulation and in SDSS observations. In observations, results can only be obtained in redshift space; these are (directly) compared with the simulation results, which are computed in real space. When examining the semi-analytical galaxies we have the benefit of omitting interlopers, thus making any comparison more qualitative than quantitative. We expect that interlopers and nuances related to the association of satellites with primaries in the observations will have the effect of weakening any inherent alignment. This will not affect our ability to quantify the dependence of the alignments on various properties of the primary or satellite galaxies.

The probabilistic nature of Bisous model allows us to define the sheets where the galactic filaments are located. The sheet normal, designated as $\mathbf{e}_1$, is well aligned with the minor principal axis of the local velocity shear (See Appendix~\ref{sec:bisous_vel}). This extends the work by \citet{Tempel:14b}, where the orientation of Bisous model filaments were compared with the major principal axis ($\mathbf{e}_3$) of the velocity shear. It implies that in observations the Bisous model can be used as an indirect tracer of velocity shear tensor in location of galactic filaments. Our main results can be summarised as following.
\begin{itemize} 
  	\item Using the SDSS observations, we found weak but statistically significant signal of alignment between the angular distribution of satellite galaxies around the isolated primary galaxies in filaments and the direction of filaments/sheets where those primaries are located. These findings confirm the predictions from simulations \citep{Libeskind:14,Forero-Romero:14}.
	\item Using the Millennium simulation populated with semi-analytical galaxies, we showed that both observations and simulations show a qualitatively compatible alignment between satellite galaxies and the orientation of filaments. These results are consistent and complement the predictions from pure $N$-body simulation \citep{Libeskind:14}.
	\item The alignment signal is much stronger for red primary galaxies and is very weak or absent for blue primaries. This trend is visible both in observations and in simulations. We also see a weak trend that the alignment signal is stronger for brighter primaries and for brighter satellite galaxies.
\end{itemize}

Our results confirm the claim that the velocity shear field dictated the infall of satellites into host systems \citep{Libeskind:14,Forero-Romero:14}. $N$-body simulations show that satellites are aligned with the principal axis corresponding to slowest collapse ($\mathbf{e}_3$) of the velocity shear and show an anti-alignment with the principal axis corresponding to fastest collapse ($\mathbf{e}_1$), whereas there are almost no alignments with $\mathbf{e}_2$ vector (intermediate principal axis). This work confirm those DM only studies using the Millennium simulation semi-analytical galaxies as well as SDSS observations.

\citet{Libeskind:14} show that the infall of satellites is universal in the sense that the strength of satellite beaming does not depend on the redshift or mass of the halo or satellite. In the present study we show that the satellite alignment depends on the colour/luminosity of the primary galaxy: brighter satellites and brighter hosts display a stronger alignment. Indeed this is in perfect agreement with the numerical studies of \citet{Libeskind:14}, which showed a similar mass dependence of the beaming effect. However, for blue/faint primaries, the satellite distribution around them may not reflect the infall direction of satellites, or the related information are disrupted when satellites fall into the main system. \citet{Agustsson:10} found the satellites distribution with respect to the major axis of primary galaxies have similar dependence on the colour of primary galaxies. However they found that the isotropic satellite distribution around blue primaries are due to that the satellite distribution dependent also on the projected distance of satellites to the primaries. The satellites close the primaries actually preferably distribute along the major axes of primaries, while those far from primaries distribute randomly or preferably along the minor axes. We also tried to explore the possible dependence of anisotropic distribution on the projected distance to the primaries for using both SDSS and model galaxies. However we found no statistical evidence supporting that. To give a conclusive answer to this question, a more detailed study based on simulations is required.

  \begin{figure}
  	\includegraphics[width=84mm]{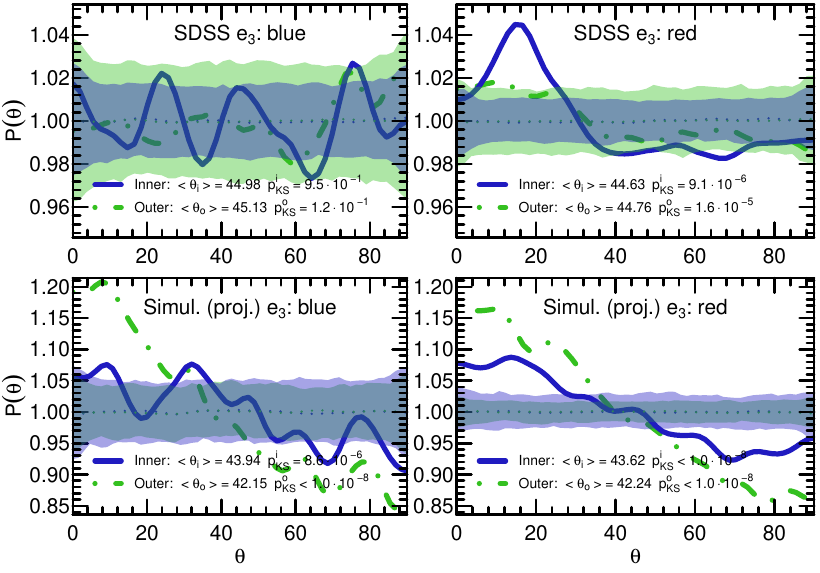}
      \caption{The alignment between anisotropic distribution of satellites and filament orientation in SDSS (upper row) and in Millennium simulation (lower row) for a fixed stellar mass ($10<\log (M_\mathrm{stellar})<11$). The alignment is shown for red (right column) and blue (left column) primary galaxies. The designation of lines and labels are the same as in Fig.~\ref{fig:sat_sdss}.}
       \label{fig:stmass}
  \end{figure}

The alignment dependence on primary colour is an interesting result and requires more attention. Logically, the colour dependence may simply reflect the mass (or luminosity) dependence of the alignment: redder primaries are also more massive and more massive galaxies show stronger alignment signal. To investigate this effect, we selected primaries on a fixed stellar mass\footnote{For SDSS stellar masses were estimated using the {\sc kcorrect} algorithm \citep{Blanton:07}. For semi-analytical galaxies, the stellar masses are direct outcome from {\sc galform}.} bin ($10<\log (M_\mathrm{stellar})<11$) and calculated the alignment signal for blue and red primaries\footnote{We also conducted the analysis using narrower stellar mass bins. The conclusions in this case remain the same, however, the significance is reduced since the number of primaries decreases.}. The resulting alignment signals for SDSS and simulations are shown in Fig.~\ref{fig:stmass}, where we see that the dependence on primary colour is intrinsic. However, the number of satellites (and significance) for blue primaries is lower than for red primaries, because red primaries have more satellites than blue ones of the same stellar mass \citep{Wang:12}.

Since redder colour typically indicate older stellar population, these results suggest that a significant alignment only exists in haloes with a relatively old stellar population. \citet{Agustsson:10} argue that satellites that recently enter to the host halo show considerably less anisotropy than do those which entered their host's halo at earlier times. Perhaps the trajectories of satellites are affected by the potential well of the surrounding filamentary environment.

Clearly, the satellite-filament correlation must hold some interesting clues regarding galaxy formation. The measured alignments between satellite positions and the surrounding filamentary environment may indicate that galaxy formation is affected by large-scale environment. The alignment signal may be natural result of how satellites accrete via streams along the direction of the filaments. The results also suggest that the red central galaxies are significantly affected by such accretion.

Recently, \citet{Lee:14b} carried out similar study as presented in this work. \citet{Lee:14b} used the SDSS DR7 observations and reconstructed the velocity shear field using linear approximation. Motivated by the work of \citet{Libeskind:14}, they show that satellites around isolated galaxies are aligned with $\mathbf{e}_3$ and show anti-alignment with $\mathbf{e}_1$. They conclude that the alignment with respect to velocity shear is universal as predicted in \citet{Libeskind:14}. In general, our study is in good agreement with the work by \citet{Lee:14b}. There are two important differences. First of all we see a dependence on primary galaxy colour and luminosity. Secondly we use a fully independent method (the Bisous model) to compute filamentary structure. The agreement with \citet{Lee:14b} supports the notion that satellites are not only beamed towards their hosts by the cosmic web, but that this effect can be seen long after the accretion process has ended.

As a complementary study, \citet{Tempel:15} shows that galaxy pairs in SDSS are also very strongly aligned with galactic filaments where they are located. Adding the results from this paper, we can conclude that not only satellite galaxies, but also galaxy pairs and potentially all structures are elongated along the galactic filaments. These results will put important constraints on the formation and evolution of galaxies and groups in the cosmic web.

\section*{Acknowledgments}

We thank the Referee for detailed report that helped to clarify many aspects in the paper. ET and RK acknowledge the support by the Estonian Ministry for Education and Science research projects IUT40-2, IUT26-2 and the Centre of Excellence of Dark Matter in (Astro)particle Physics and Cosmology (TK120).
NIL and QG acknowledges the DFG for support in completing this project.
Funding for SDSS-III has been provided by the Alfred P. Sloan Foundation, the Participating Institutions, the National Science Foundation, and the U.S. Department of Energy Office of Science. The SDSS-III web site is http://www.sdss3.org/.
SDSS-III is managed by the Astrophysical Research Consortium for the Participating Institutions of the SDSS-III Collaboration including the University of Arizona, the Brazilian Participation Group, Brookhaven National Laboratory, Carnegie Mellon University, University of Florida, the French Participation Group, the German Participation Group, Harvard University, the Instituto de Astrofisica de Canarias, the Michigan State/Notre Dame/JINA Participation Group, Johns Hopkins University, Lawrence Berkeley National Laboratory, Max Planck Institute for Astrophysics, Max Planck Institute for Extraterrestrial Physics, New Mexico State University, New York University, Ohio State University, Pennsylvania State University, University of Portsmouth, Princeton University, the Spanish Participation Group, University of Tokyo, University of Utah, Vanderbilt University, University of Virginia, University of Washington, and Yale University.


\appendix

\section{Correlation between velocity shear tensor and Bisous model filaments and sheets}
\label{sec:bisous_vel}

\begin{figure*}
	\includegraphics[width=176mm]{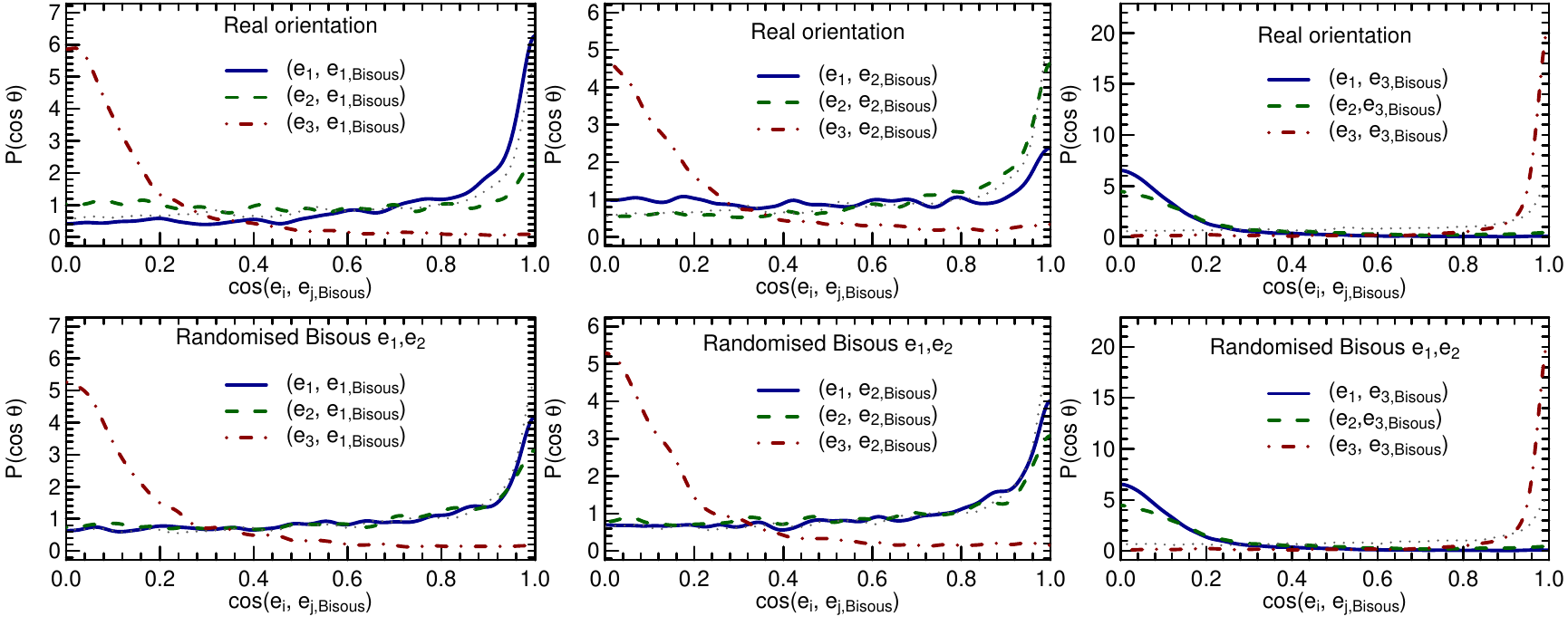}
    \caption{The alignment between Bisous model filaments and sheets ($\mathbf{e}_{i,\mathrm{Bisous}}$ vectors) and the underlying velocity field as defined by the principal axes of velocity shear tensor ($\mathbf{e}_i$ vectors). Left panel shows the alignment between $\mathbf{e}_{1,\mathrm{Bisous}}$ and $\mathbf{e}_i$; middle panel between $\mathbf{e}_{2,\mathrm{Bisous}}$ and $\mathbf{e}_i$; right panel between $\mathbf{e}_{3,\mathrm{Bisous}}$ and $\mathbf{e}_i$. The vectors $\mathbf{e}_{3,\mathrm{Bisous}}$ and $\mathbf{e}_3$ define the orientation of filaments, while $\mathbf{e}_{1,\mathrm{Bisous}}$ and $\mathbf{e}_1$ define the sheet normals. With dotted grey lines we show the random distribution between $\mathbf{e}_{1,2,\mathrm{Bisous}}$ and $\mathbf{e}_{2,1}$ assuming perfect alignment between $\mathbf{e}_{3,\mathrm{Bisous}}$ and $\mathbf{e}_3$. Strong alignment between the same principal axes of Bisous structures and shear tensor is visible, hence, the Bisous model structures trace the underlying velocity field. In lower row, the orientations of $\mathbf{e}_{1,2,\mathrm{Bisous}}$ are randomised, while keeping the $\mathbf{e}_{3,\mathrm{Bisous}}$ orientation. Comparing with upper row we note that the real orientation of $\mathbf{e}_{1,\mathrm{Bisous}}$ and $\mathbf{e}_{2,\mathrm{Bisous}}$ is not drawn from a random distribution.}
     \label{fig:Bisous_vel}
\end{figure*}

Here we analyse the correlation between velocity field (shear tensor) and the structures defined in the Bisous model using a pure DM $N$-body simulation. We extend the work presented in \citet{Tempel:14b}, that showed that the direction of filaments defined using point processes applied to the halo distribution in DM $N$-body simulation matches very well the eigenvector of the velocity shear tensor corresponding to slowest collapse. In \citet{Tempel:14b} only the orientation of filaments were compared with the $\mathbf{e}_3$ vector of the local velocity shear. Now we employ the probabilistic nature of the Bisous process and define the galactic sheets (their normals) in the location of filaments (see Sect.~\ref{sec:bisous_b1}). The aim is to test whether the halo/galaxy distribution in filaments can be used as a tracer of velocity shear tensor. For that the sheet normal defined in the Bisous model are compared with the eigenvectors of the local velocity shear tensor.

\subsection{$N$-body simulation and the velocity shear tensor}

The simulation we use in this Appendix is exactly the same as used in \citet{Tempel:14b}. For convenience a brief description is given below.

A DM-only $N$-body simulation is run assuming the standard $\Lambda$CDM concordance cosmology, in particular a flat universe with cosmological constant density parameter $\Omega_\Lambda= 0.72$, matter density parameter $\Omega_\mathrm{m} = 0.28$, a Hubble constant parameterized by $H_0 = 100~h~\mathrm{km}~\mathrm{s}^{-1}\mathrm{Mpc}^{-1}$ (with $h = 0.7$), a spectral index of primordial density fluctuations given by $n_s = 0.96$, and mass fluctuations given by $\sigma_8 = 0.817$.

The simulations span a box of side length $64~h^{-1}\mathrm{Mpc}$ with $1024^3$ particles, achieving a mass resolution of $\sim\!1.89 \times 10^7~h^{-1}M_\odot$ and a spatial resolution of $1~h^{-1}\mathrm{kpc}$. The publicly available halo finder AHF \citep{Knollmann:09} is run on the particle distribution to obtain a halo catalogue. Only haloes more massive than $10^9~h^{-1}M_\odot$ are considered in this work.

We quantify the cosmic web by means of the velocity shear tensor. This method is described in detail in \citet{Hoffman:12}. The salient aspects are highlighted here, in brief. The cosmic velocity field is calculated using a ``Clouds-in-Cell'' (CIC) algorithm on a $256^{3}$ grid. The velocity (and density) fields are then smoothed with a Gaussian kernel equal to at least one cell (i.e. $r_{\rm smooth}=0.25~h^{-1}$Mpc) in order to get rid of the spurious artificial cartesian grid introduced by the CIC. In practice the smoothing sets the scale of the calculation. In current study, we use only the smoothing scale 0.5~$h^{-1}$Mpc. The velocity shear tensor is defined as $\Sigma_{\alpha\beta}=\frac{1}{2H_{0}}\big(\frac{\partial v_{\alpha}}{\partial r_{\beta}}+\frac{\partial v_{\beta}}{\partial r_{\alpha}}$\big) and is calculated by means of fast Fourier transform (FFT) in $k$-space. The velocity shear tensor is then diagonalized and its eigenvectors and eigenvalues are identified. We denote the eigenvectors as $\mathbf{e}_1$, $\mathbf{e}_2$ and $\mathbf{e}_3$, where $\mathbf{e}_1$ is the direction of fastest collapse and $\mathbf{e}_3$ gives the direction of slowest collapse (e.g. the direction of filaments).

Note that the velocity shear field is identical to the tidal field, defined as the Hessian of the potential, namely $T_{\alpha\beta}=\frac{\partial^{2}\phi}{\partial r_{\alpha}\partial r_{\beta}}$, when smoothed on large enough (i.e. $>$ few Mpc) scales.

\subsection{Orientation of Bisous filaments and sheets with respect to the underlying velocity field}

\citet{Tempel:14b} showed that the Bisous filaments are very well aligned with the slowest collapse in underlying velocity field. Here, we extend the study and analyse the orientation of Bisous model filaments and sheets with respect to the all three eigenvectors of the velocity shear tensor. For that we calculate the distribution of cosine of the angles between the $\mathbf{e}_{i,\mathrm{Bisous}}$ (Bisous model) and $\mathbf{e}_i$ (velocity shear) vectors. The angle between Bisous structures and velocity field is found in the location of filament axes. For every point in the filament spine, the angle is found between the orientation of that filament point ($\mathbf{e}_{i,\mathrm{Bisous}}$) and the velocity shear eigenvector ($\mathbf{e}_i$) defined in the closest cell. Since the scale and spacing of points along the filament spine is $0.5~h^{-1}\mathrm{Mpc}$, which is larger than the cell size ($0.25~h^{-1}\mathrm{Mpc}$), interpolation of velocity shear to the filament axis is not required.

Fig.~\ref{fig:Bisous_vel} shows the alignment between the structure orientations in the Bisous model ($\mathbf{e}_{1,\mathrm{Bisous}}$, $\mathbf{e}_{2,\mathrm{Bisous}}$, $\mathbf{e}_{3,\mathrm{Bisous}}$) and the velocity shear ($\mathbf{e}_1$, $\mathbf{e}_2$, $\mathbf{e}_3$). Fig.~\ref{fig:Bisous_vel} shows that in addition to the very strong alignment between $\mathbf{e}_{3,\mathrm{Bisous}}$ and $\mathbf{e}_3$, there is also a strong alignment between $\mathbf{e}_{1,\mathrm{Bisous}}$ and $\mathbf{e}_1$, and between $\mathbf{e}_{2,\mathrm{Bisous}}$ and $\mathbf{e}_2$. This shows that the sheet normal $\mathbf{e}_{1,\mathrm{Bisous}}$ defined in the location of Bisous filaments is aligned with the fastest collapse in the underlying velocity field, e.g. the Bisous model can be used as an indirect tracer of underlying velocity field in cases where the velocity information is not available (e.g. in observations).

It is expected that if $\mathbf{e}_{3,\mathrm{Bisous}}$ and $\mathbf{e}_3$ are perfectly aligned, then even for completely random distribution for $\mathbf{e}_{1,\mathrm{Bisous}}$, the latter is somewhat aligned with $\mathbf{e}_1$ and $\mathbf{e}_2$ in real space. To show that this is not so in Bisous structures, we randomised the $\mathbf{e}_{1,\mathrm{Bisous}}$ while keeping the $\mathbf{e}_{3,\mathrm{Bisous}}$ orientation (direction of filament). The alignments for randomised  $\mathbf{e}_{1,2,\mathrm{Bisous}}$ are shown in lower panel of Fig.~\ref{fig:Bisous_vel}. If we compare the alignments in upper and lower panels in Fig.~\ref{fig:Bisous_vel}, we clearly see that the Bisous model $\mathbf{e}_{1,2,\mathrm{Bisous}}$ vectors are not drawn from random distribution. The alignment between the sheet normal in Bisous model ($\mathbf{e}_{1,\mathrm{Bisous}}$) and velocity field eigenvector $\mathbf{e}_1$ is real.

\section{Known alignments of satellites and primaries}

  \begin{figure}
	\includegraphics[width=84mm]{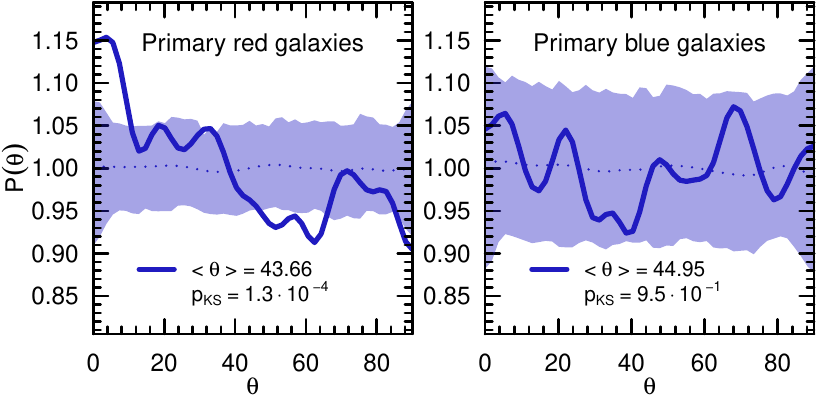}
    \caption{The excess probability of primaries as a function of angles
     between the major axis of primaries and the filament axes. The results
     for red and blue primaries are shown in left and right panel, respectively.
     Dotted lines with filled region show the null-hypothesis together with its
     95\% confidence limit.
    \label{fig:prim_fil}} 
  \end{figure}
  
  \begin{figure}
	 \includegraphics[width=84mm]{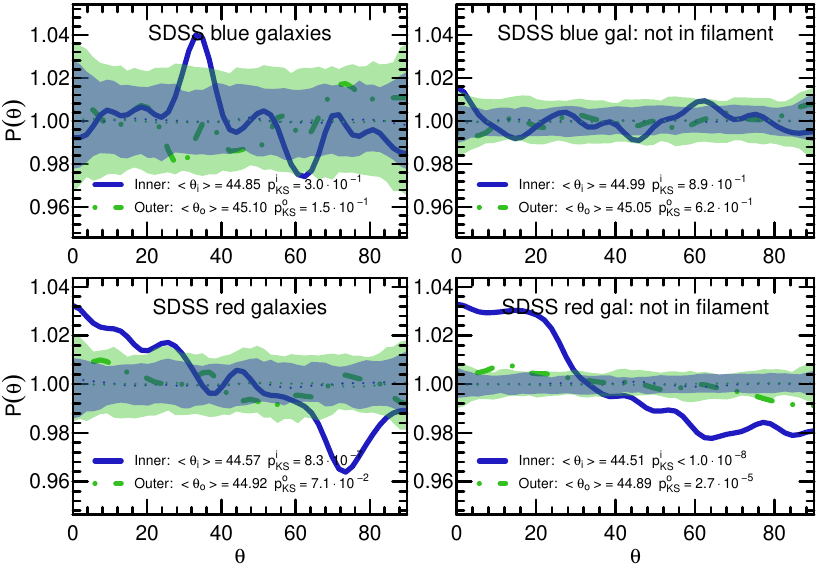}
    \caption{The normalised number of satellites as a function of the angle 
    between the satellite-primary position vector and the main axis of the
    primary galaxies (position angle $\phi$) in sky plane. The alignment is shown for blue (top row) and red (bottom row) primary galaxies, and for primaries in filaments (left column) and not in filaments (right column).}
    \label{fig:prim_sat}
  \end{figure}

  \subsection{The alignment between the major axis of primaries and filament axes}
  \label{sec:pf}
  
    In this section, we corroborate that the major axis of primaries aligns with the filament axes, which has already been studied \citep[e.g.][]{Tempel:13a}. Fig.~\ref{fig:prim_fil} shows the excess probability of primaries as a function of the angle between the major axis of primaries and filaments axes. The results corroborate that for red primaries, the major axis preferentially align with the filaments axes. Here we used the de~Vaucouleurs fit position angle ($\phi$) as the proxy of the major axis of the primaries.

  \subsection{The alignment between anisotropic distribution of satellites and  the
  major axis of primaries}
    \label{sec:ps}

    In this section, we corroborate that anisotropic distribution of satellites align with the major axis of primaries both in-filaments and not-in filaments. The result are shown in Fig.~\ref{fig:prim_sat}, which indicate that for primaries both in-filaments and not-in-filaments, the anisotropic distribution of satellites align with the major axis of red ones, but randomly distributed with respect to the the major axes of blue ones. It is consistent with the results from other studies \citep[e.g.][]{Yang:06}.

\label{lastpage}

\end{document}